\documentclass[preprint,showpacs,prb,floatfix,superscriptaddress,citeautoscript,cite]{revtex4}
\usepackage{graphicx}
\usepackage{color}
\usepackage{amsmath}

\begin{document}

\title{Stable Monolayer $\alpha$-Phase of CdTe: Strain-Dependent Properties}

\author{E. Unsal}
\email{elifunsal@iyte.edu.tr}
\affiliation{Department of Physics, Izmir Institute of Technology, 35430, Izmir, 
Turkey}

\author{R. T. Senger}
\affiliation{Department of Physics, Izmir Institute of Technology, 35430, Izmir, Turkey}
\affiliation{ICTP-ECAR Eurasian Center for Advanced Research, Izmir Institute of Technology, 35430, Izmir, Turkey.}

\author{H. Sahin}
\affiliation{ICTP-ECAR Eurasian Center for Advanced Research, Izmir Institute of Technology, 35430, Izmir, Turkey.}
\affiliation{Department of Photonics, Izmir Institute of Technology, 35430 Izmir, Turkey}

\date{\today}


\begin{abstract}
CdTe is a well known and widely used binary compound for optoelectronic applications. In this study, we propose the thinnest, free standing monolayer of CdTe which holds the tetragonal-PbO ($\alpha$-PbO) symmetry. The structural, electronic, vibrational and strain dependent properties are investigated by means of first principles calculations based on density functional theory. Our results demonstrate that the monolayer $\alpha$-CdTe is a dynamically stable and mechanically flexible material. It is found that the thinnest monolayer crystal of CdTe is a semiconductor with a direct band gap of 1.95 eV, which corresponds to red light in the visible spectrum. Moreover, it is found that the band gap can be tunable under biaxial strain. With its strain-controllable direct band gap within the visible spectrum, stable $\alpha$-phase of monolayer CdTe is a suitable candidate for optoelectronic device applications. 
\end{abstract}

\maketitle

\section{Introduction}

Since the technology proceeds towards nanoscale, synthesis of low-dimensional materials has become more of an issue. Graphene, one-atom-thick crystal of C-atoms arranged in honeycomb structure, has been successfully synthesized in 2004. \cite{Novoselov1} Although graphene possesses extraordinary physical properties, \cite{Novoselov2,Geim} having a zero band gap in its electronic structure restricts its application in nanotechnology. \cite{Zhang} Over the last decade, graphene has also been a pioneer in exploring novel two-dimensional (2D) materials in various categories such as transition metal dichalcogenides \cite{Fang,Ali,Keum} (e.g. MoS$_2$ \cite{Mak,Kong,Radisavljevic,Kang1,Esfahani} and WS$_2$ \cite{Mahler,Elias} ), and II-VI binary compounds (e.g. CdSe  \cite{Son} and ZnSe \cite{Sun,Park} ). The II-VI group semiconductors are well-known materials and there have been wide range of theoretical and experimental studies on these materials. \cite{Li1,Qu,Huang,Hines,Manna} 

As a II-VI binary compound, bulk CdTe was widely studied in the last half-century \cite{Murray,Vogel,Zakharov,Ley,Chelikowsky}. The bulk CdTe has a direct band gap of approximately 1.50 eV, which is optimally matched to the solar spectrum. \cite{Wu} Therefore, this material has an extensive usage in optoelectronic device applications such as photo-detectors, infrared and gamma-ray detectors. \cite{Pennicard,Kanevce,Gupta,Rogalski,Hibino} Beyond its bulk form, dimensionally reduced CdTe structures, e.g. quantum dots and rods \cite{PengPeng}, have attracted great attention due to their composition- and size-dependent absorption and emission spectrum. \cite{Alivisator,Michalet} Gupta \textit{et al.} succeeded in reducing the thickness of CdTe film in CdS/CdTe solar cell, which is the thinnest CdTe cell with high efficiency.\cite{Gupta2} Moreover, Sun \textit{et al.} stated that they achieved to improve quantum-dot-based light emitted diodes (LEDs) with enhanced electroluminescent efficiency and lifetimes for long-term operations. In consideration of their findings, they also reported that the quantum-dot-based LEDs will be used in manufacture of flat-panel displays (such as televisions and monitors). \cite{nanoSun} Recently, Ithurria \textit{et al.} reported the synthesis of cadmium chalcogenide (CdSe, CdS and CdTe) nanoplatelets in various thickness. They showed that these class of colloidal semiconductor materials exhibit physical and optical properties such as tunable thickness, controllable lateral dimension and enhanced oscillator strength, which makes the colloidal platelets suitable for nonlinear optical devices.\cite{Ithurria1} In spite of intensive studies on low dimensional CdTe, monolayer phase of CdTe is never investigated before.

Herein, we investigate structural, electronic and mechanical properties of a phase of monolayer CdTe by employing the first principles calculations based on DFT. The monolayer CdTe is found to be dynamically stable and has two prominent Raman-active modes. The monolayer CdTe is a direct gap semiconductor with its strain-tunable band gap energy and has a low in-plane stiffness.

The paper is organized as follows: in Sec. \ref{computational} information about our computational methodology is given. Then, possible structural phases of monolayer CdTe are discussed in Sec. \ref{structural-phases}. Identification of 
the structural phase of the monolayer $\alpha$-CdTe is presented in Sec. 
\ref{structure}. In Sec. \ref{dynamical} the vibrational properties are presented. The electronic properties of the structure are given in the Sec. \ref{electronic} and the strain dependent properties are declared in Sec. \ref{strain}. Final section, Sec. \ref{conclusion}, is allocated for the conclusion of our study.

\begin{figure*}
\includegraphics[width=16 cm]{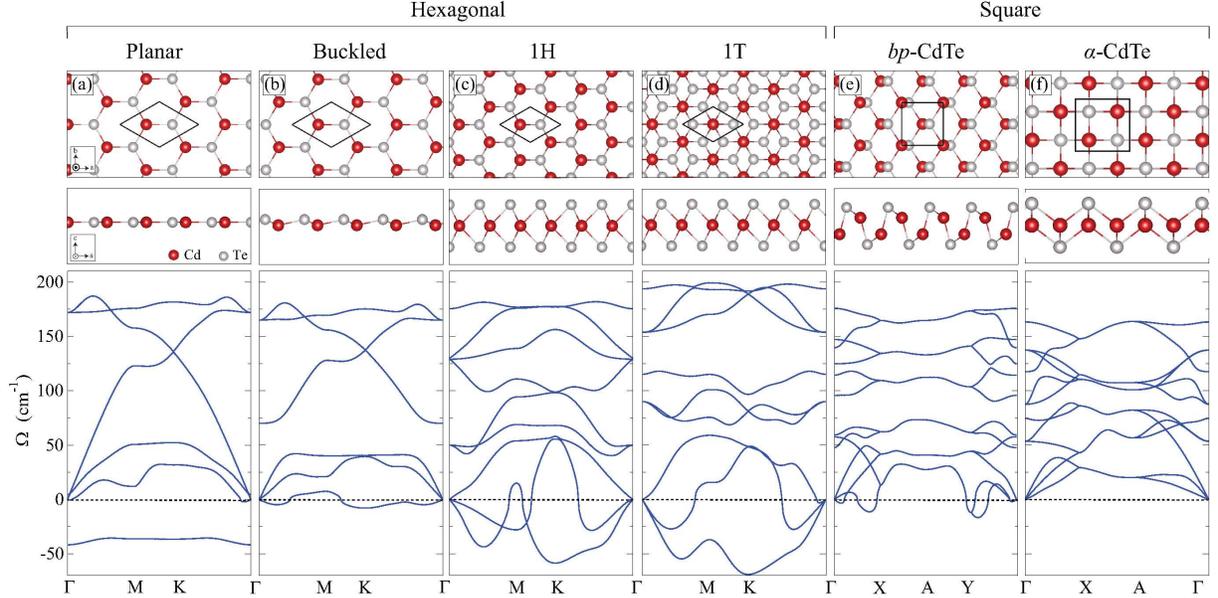}
\caption{\label{phases}
(Color online) Top and side views of various monolayer CdTe phases and their phonon band diagrams.   }
\end{figure*}

\section{Computational Methodology}\label{computational}

All calculations were performed by utilizing DFT and projector-augmented wave (PAW) potentials as implemented in the Vienna \textit{ab-initio} simulation package (VASP). \cite{vasp1,vasp2,vasp3} Perdew-Burke-Ernzerhof (PBE) form of generalized gradient approximation (GGA) \cite{GGA-PBE} was used for the description of electron exchange and correlation. The Heyd-Scuseria-Ernzerhof (HSE06) hybrid functional was included on top of GGA for band gap estimation.\cite{HSE06} The van der Waals forces were included by using the DFT-D2 method of Grimme.\cite{vdW1,vdW2} Bader technique was adopted in order to determine the charge transfer in the structure.\cite{Bader1,Bader2} 

The structure was relaxed until the total energy difference between consecutive electronic steps reached at the level of 10$^{-5}$ eV and Hellmann-Feynman forces on each unit cell was less than 10$^{-4}$ eV/\AA{}. Gaussian smearing was used with a broadening of 0.05 eV and the pressure on the unit cell was reduced to a value less than 1.0 kB in all three directions. To avoid interactions between adjacent monolayers, $∼$12 \AA{} vacuum space was included. For the Brillouin Zone (BZ) integration, $16\times16\times1$ $\Gamma$-centered mesh was used for the primitive unit cell. 

Moreover, the vibrational properties of monolayer $\alpha$-CdTe were investigated for  $6\times6\times1$ supercell with 144 atoms by using the small displacement method as implemented in PHON code. \cite{Alfe} For the calculation of cohesive energy per atom, $E_{Cohesive}$, we used the following formula,
\begin{equation}
E_{Cohesive}= \frac{1}{n_{tot}} \left[n_{Cd}E_{Cd} + n_{Te}E_{Te} - E_{ML} \right],
\label{eq:coh}
\end{equation}
where  E$_{Cd}$ and E$_{Te}$ represent the energies of single isolated Cd and Te atoms. E$_{ML}$ stands for the total energy of the monolayer $\alpha$-CdTe and $n_{tot}$, $n_{Cd}$ and $n_{Te}$ denote the total number of atoms, number of Cd and Te atoms within in the unit cell, respectively. Calculated values of the cohesive energies are discussed in Sec. \ref{structural-phases}.

For the calculation of the elastic constants of $\alpha$-phase, $4\times4\times1$ 64-atom supercell was considered. Strains were applied along $x$- and $y$-axis by changing the lattice parameters along each direction. We applied uniaxial strains, $\varepsilon_x$ and $\varepsilon_y$, and biaxial strain along both directions. Strain energy $E_S$ was calculated by subtracting the ground state energy of the system from the energy of the system under applied load. By quadratic regression, calculated data was fitted to following equation,
\begin{equation}
E_S=c_1\varepsilon_x ^2 + c_2\varepsilon_y ^2 + c_3\varepsilon_x\varepsilon_y
\label{eq:Es} 
\end{equation}
and the coefficients of $c_i$ were obtained. In-plane stiffness along $x$- and $y$-axis is calculated by using the following formulas,\cite{Topsakal} 
\begin{align}
 C_x= \frac{1}{A_0} \left( 2c_1- \frac{ c_{3}^{2}}{2c_2} \right)  \label{eq:C1} \\
 C_y= \frac{1}{A_0} \left( 2c_2- \frac{ c_{3}^{2}}{2c_1} \right) 
 \label{eq:C2}
\end{align}
where $A_0$ is the unit cell area of the unstrained system and constants of $c_1$, $c_2$ and $c_3$ attained from Eq. \ref{eq:Es}. For the calculation of Poisson's ratio, we benefited from $\nu_x=c_{3}/2c_1$ and $\nu_y=c_{3}/2c_2$ formulas, where $\nu_x$ and $\nu_y$ are the ratios in the directions of $x$ and $y$. \cite{Topsakal} The calculated values of the elastic constants are discussed in Sec. \ref{strain}.

\section{Structural Phases of Monolayer C\MakeLowercase{d}T\MakeLowercase{e} }\label{structural-phases}
 
\begin{figure}
\includegraphics[width=10 cm]{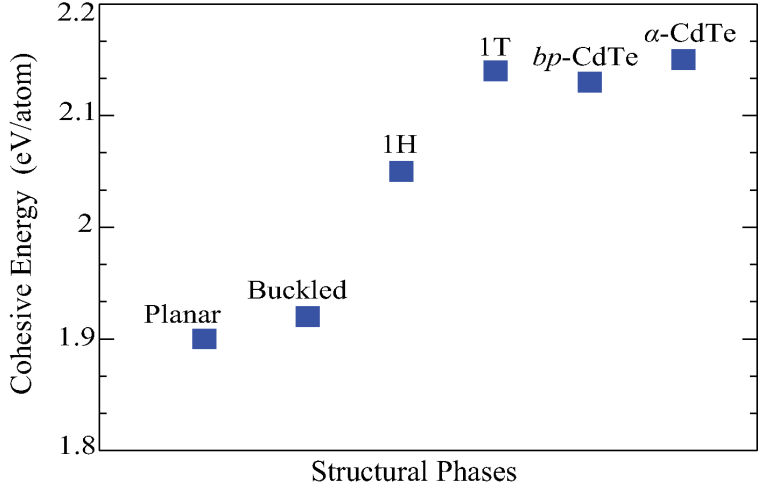}
\caption{\label{plot}
(Color online) Calculated cohesive energy values of possible phases of monolayer CdTe are demonstrated. $\alpha$-CdTe is energetically the most favorable structure among these phases.  }
\end{figure}

In this section, we investigate two-dimensional phases of CdTe, which include planar hexagonal (graphene-like), buckled hexagonal (silicene-like), 1T and 1H phases. As square lattices, black phosphorus ($bp$-CdTe) and $\alpha$-CdTe structural phases are also studied. In order to find out the energetically the most favorable phase for the monolayer CdTe, we calculate the cohesive energy of each phase and compare with each other. We also perform the phonon calculation through the whole BZ for each phase to analyze their dynamical stability. It is found that the cohesive energies for the 1H, planar and buckled hexagonal phases are found to be 2.05 eV/atom, 1.90 and 1.92 eV/atom, respectively. As seen in Fig. \ref{plot}, planar hexagonal structure is energetically the least favorable phase among all of these structures and the phonon-band diagram of the planer phase reveals its dynamical instability. The cohesive energies of the monolayer $bp$-CdTe, 1T and $\alpha$-CdTe phases are calculated to be 2.13, 2.14 and 2.15 eV/atom, respectively. Although these energy values are very close to each other, it is seen that 1T and $bp$-CdTe phases are dynamically unstable. As seen in Fig. \ref{phases}, all of the hexagonal structures as well as $bp$-CdTe have negative frequencies.

\begin{figure}
\includegraphics[width=10 cm]{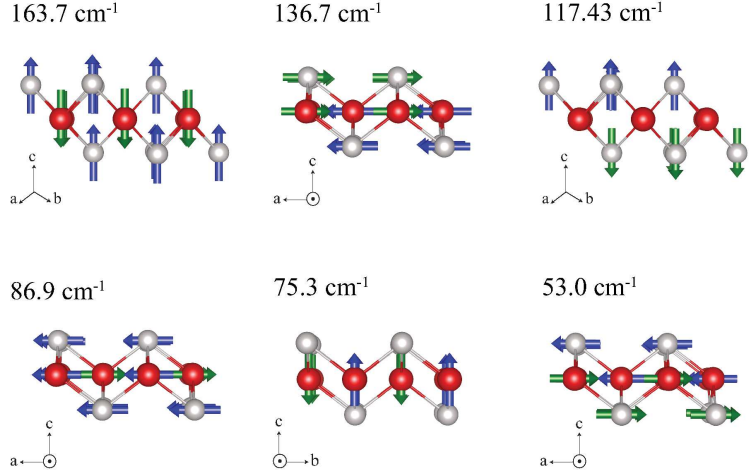}
\caption{\label{eigen}
(Color online) Phonon modes at $\Gamma$ point for monolayer $\alpha$-CdTe. Red and grey atoms represent the Cd and Te atoms, respectively.}
\end{figure}

Among these monolayer phases of CdTe, $\alpha$-phase is energetically the most favorable structure, which indicates that the monolayer CdTe is more likely to have the $\alpha$-phase in comparison to other possible structural phases. In addition, the $\alpha$-phase is found to be dynamically stable. Therefore, in the following sections, characteristic properties of the monolayer $\alpha$-CdTe are presented in detail.

\section{Structural Properties of $\alpha$-C\MakeLowercase{d}T\MakeLowercase{e}}\label{structure}

As a well-known semiconducting material, CdTe crystallizes in the zinc-blende (\textit{zb-}) structure under ambient conditions of temperature and pressure in its bulk form. By varying the pressure, there can occur structural phase transition and it is possible to observe rocksalt, cinnabar or orthorhombic phases of bulk CdTe. \cite{Madelung} Here, we investigate $\alpha$-PbO-type monolayer CdTe, in which a planar layer of square Cd lattice is bonded to Te atoms tetrahedrally as seen in Fig. \ref{phases}. These type of structures belong to the space group of \textit{P4/nmm}. \cite{spacegroup,spacegroup2,spacegroup3} Primitive cell is square and includes two Cd and two Te atoms. As seen in Table \ref{table1}, the lattice parameter of the monolayer $\alpha$-CdTe is calculated to be $∼$4.66 \AA{}, which is smaller than that of bulk \textit{zb}-CdTe, and the thickness (vertical distance between uppermost and lowermost Te atoms) is found to be  $∼$3.55 \AA{}. All the Cd-Te bonds in the monolayer are found to be equal with a value of $∼$2.93 \AA{}, larger than that of the $zb$-bulk, thus it is expected that the Cd-Te bonds in bulk structure is quite stronger. Bader charge analysis reveals that each Cd atom donates $0.5e$ to each Te atom which indicates a polar-covalent type bond. Moreover, we calculate the cohesive energies of the monolayer and $zb$-bulk structure in order to understand the formation of the monolayer structure. The cohesive energy is found to be 2.15 and 2.32 eV/atom for the monolayer and the bulk $zb$-CdTe, respectively. Ward \textit{et al.}\cite{Ward1} reported that the different phases of the CdTe can exist with various cohesive energies in the range between $\sim$1.75-2.20 eV/atom. However, the obtained cohesive energy values vary depending on the used DFT methodology.\cite{Zhou2,Ward2}

\begin{table*}
\caption{\label{table1} Calculated parameters for monolayer $\alpha$-CdTe are; the lattice constant, \textit{a}; the atomic distance between Cd and Te atoms, d$_{Cd-Te}$; the charge donation of Cd, $\Delta\rho$; the cohesive energy per atom, E$_{Cohesive}$. E$^{PBE}_{g}$, E$^{PBE+SOC}_{g}$ and E$^{PBE+SOC+HSE06}_{g}$ are the energy gap values calculated with PBE, PBE+SOC and PBE+SOC with HSE06 method, respectively. $\Phi$ and $\mu$ present work function and magnetization, respectively. For the calculations of zb-bulk,  conventional cubic cell which consists of eight atoms is considered and work function values are taken from Ref.\cite{workfunction-bulk}}
\begin{tabular}{rccccccccccccc}
\hline\hline
 & \textit{a} &  d$_{Cd-Te}$ &  $\Delta\rho$ &  E$_{Cohesive}$ &  E$^{PBE}_{g}$ & E$^{PBE+SOC}_{g}$ & E$^{PBE+SOC+HSE06}_{g}$ & $\Phi$ & $\mu$ \\
 & (\AA)  & (\AA) &  ($e^{-}$) & (eV/atom)  & (eV) & (eV)   & (eV) & (eV) &($\mu_{B}$) \\
\hline
monolayer          & 4.66   & 2.93 & 0.5 &  2.15   & 1.28   & 1.02 & 1.95  & 5.20      & 0 \\
\textit{zb}-bulk   & 6.52   & 2.82 & 0.5 &  2.32   & 0.72   & 0.45 & 1.25  & 5.40 $\sim$ 5.65  & 0 \\
\hline\hline
\end{tabular}
\end{table*}

\section{Phonons: Dynamical Stability}\label{dynamical}

In order to investigate the vibrational properties of monolayer $\alpha$-CdTe, the atoms are slightly distorted from their equilibrium positions. It is found that $\alpha$-CdTe crystal can generate the required restoring force and remains dynamically stable. As it is seen in Fig.\ref{phases} (f), the vibrations occur at low frequencies smaller than 200 cm$^{-1}$, which indicate the soft phonon modes, i.e. flexibility of the structure. Since Cd and Te atoms are relatively large atoms, Cd-Te bond length in the crystal is large which means that the $\alpha$-CdTe has a flexible nature.

\begin{figure}
\includegraphics[width=15 cm]{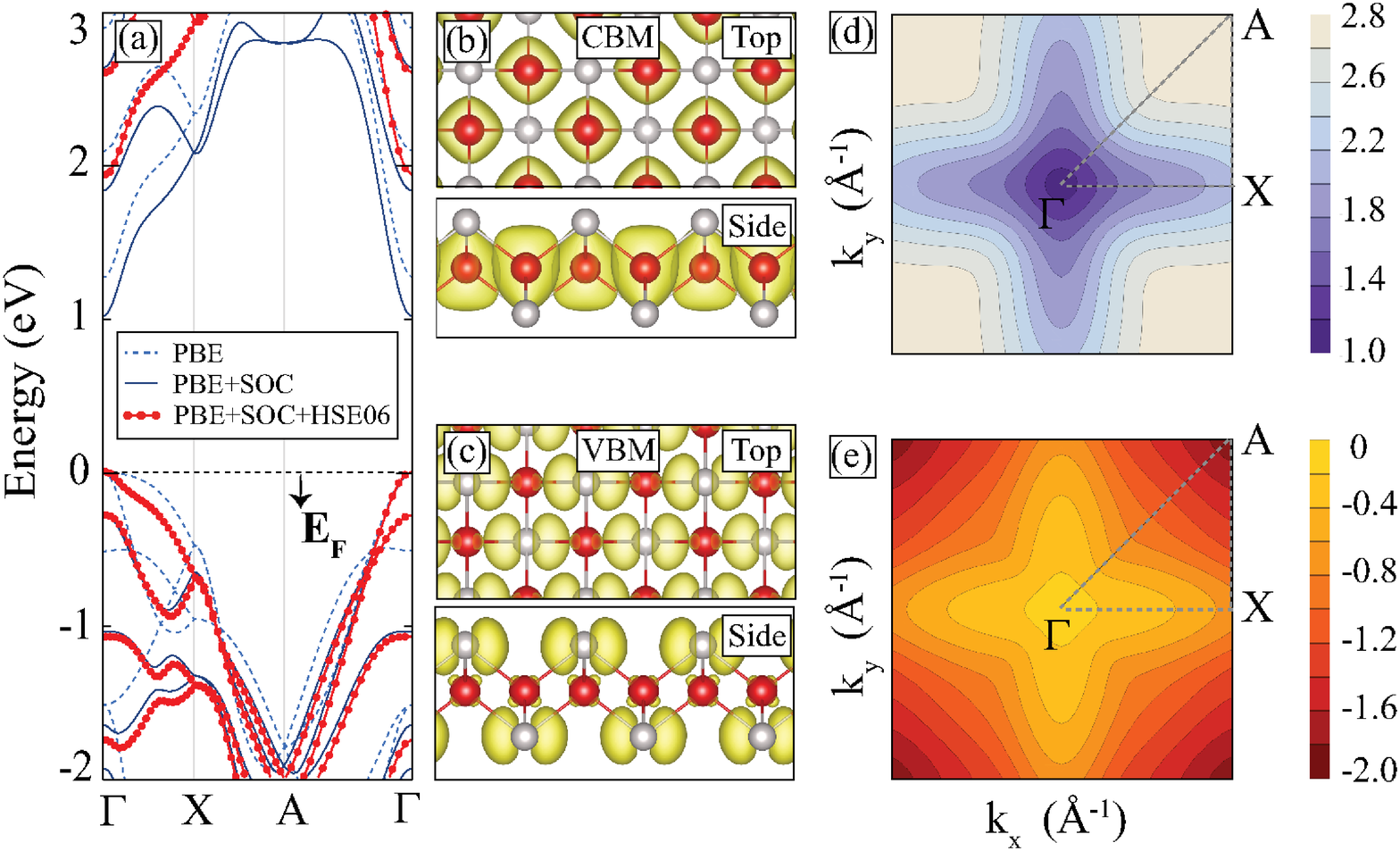}
\caption{\label{elec}
(Color online) (a) The energy-band structure of monolayer $\alpha$-CdTe. Charge densities of (b) CBM and (c) VBM are demonstrated. 2D-contour plots of (d) CBM and (e) VBM are drawn with data obtained by PBE+SOC calculation. The dashed line represent the irreducible BZ. The color scale demonstrates the energy values in unit of eV and the Fermi level ($E_F$) is set to zero.  }
\end{figure}

The phonon spectrum of the monolayer $\alpha$-CdTe includes 12 phonon branches. 3 of which are the acoustic (longitudinal acoustic (LA), transverse acoustic (TA) and out-of-plane flexural (ZA)), and 9 of them are the optical vibrational modes. 6 of the optical vibrational modes are demonstrated in Fig.\ref{eigen} due to the degeneracy in three of them. The optical vibrational mode with the highest frequency, which is found at 163.7 cm$^{-1}$, corresponds to an out-of-plane mode in which both of Te sub-layers move opposite to the Cd sub-layer. Moreover, there are three doubly degenerate modes at 53.0, 86.9 and 136.7 cm$^{-1}$, respectively. The character of the modes are in-plane and at all frequencies, Cd atoms move opposite to each other. In the phonon branches at 53.0 and 136.7 cm$^{-1}$, the top and the bottom Te sub-layers move in opposite directions; however, both of the Te sub-layers move in same direction at 86.9 cm$^{-1}$.

Moreover, Raman-intensity calculations are performed for the monolayer $\alpha$-CdTe and it is found that the structure has two prominent Raman-active modes with frequencies of 117.4 and 75.3 cm$^{-1}$. Both of the Raman-active modes are singly degenerate and have characteristics of out-of-plane counter-phase motion. At frequency of 75.3 cm$^{-1}$, Te atoms remain stationary, while Cd atoms move in opposite directions with respect to each other. At frequency of 117.4 cm$^{-1}$, Cd atoms are immobile and Te sub-layers move in opposite directions with respect to each other. Raman activity of these two modes is expected due to the in-plane inversion symmetry of their motions.

In addition, the structures, found to be individually stable, become stable on the substrates as well. We search the suitable surfaces that the monolayer $\alpha$-CdTe can grow on. Thus, the substrate of $\alpha$-FeSe with same structural symmetry as $\alpha$-CdTe  were investigated first. For instance, $\alpha$-FeSe thin films, belong to \textit{P4/nmm} space group as $\alpha$-CdTe, can grow on (001) surface of GaAs substrates by using low-pressure metal organic vapor deposition technique. \cite{Liu} More recently, it is observed that single-layer $\alpha$-FeSe films can grow on SrTiO$_3$ perovskite. \cite{Ge} As members of cubic perovskite group, RbCaF$_3$, CsCaF$_3$ and CsIO$_3$ have lattice constants of 4.45 \AA{}, 4.52 \AA{} and 4.67 \AA{}, \cite{Verma} respectively. Since CsIO$_3$ has lattice constant very close to that of $\alpha$-CdTe, it can be an ideal substrate for growth of CdTe monolayer. 

\section{Electronic Properties of $\alpha$-C\MakeLowercase{d}T\MakeLowercase{e}}\label{electronic}

The $zb$-CdTe is a well-known direct band gap semiconductor with a band gap ranging from 1.37 to 1.54 eV. \cite{Fonthal} In this section, we examine the electronic structure of the monolayer $\alpha$-CdTe and compare its properties with that of $zb$-structure.

Firstly, the band diagram of the monolayer $\alpha$-CdTe is calculated by using PBE functional and an energy gap value of 1.28 eV is found as seen in Table \ref{table1}. Our calculations reveal that it has a direct band gap at symmetry point of $\Gamma$. Inclusion of the spin-orbit coupling (SOC) leads to splitting of 0.28 eV at $\Gamma$ point and a decrease in the energy gap values as seen in Fig. \ref{elec} (a). In order to obtain accurate band gap values, HSE06 method is included on top of SOC and the results show that the band gap values of both structures are increased as seen in Table \ref{table1}. It is found that $zb$-bulk and the monolayer $\alpha$-CdTe structures have energy gap values of 1.25 eV and 1.95 eV, respectively.

Moreover, for valance band maximum (VBM) and conduction band minimum (CBM), 2D-contour plots are drawn for the SOC-added calculations by taking the first BZ as a limit. As seen in Fig. \ref{elec} (e), adjacent colors indicate the changes in energy of 0.2 eV and the energy values for VBM at points of $\Gamma$, $X$ and $A$ are 0, -0.6 and -2.0 eV, respectively. In addition, VBM and CBM are visualized in 3D as seen in Fig.\ref{elec} (d).

We also investigate the characteristics of the VBM and the CBM and calculate the band decomposed charge densities for the monolayer $\alpha$-CdTe. The results indicate that the VBM is mostly possessed of the $p_x$- and $p_y$-orbitals of the Te atoms; however, there is small contribution of $d_{yz}$- and $d_{xz}$-orbitals of Cd atoms. The CBM is dominated by the $s$-orbitals of the Cd atoms.

Effective mass, a unique feature of the material, is inversely proportional to the conductivity. In order to analyze conduction properties of $\alpha$-CdTe, effective mass of electrons and holes is calculated. In the directions of $\Gamma \rightarrow X$ and $\Gamma \rightarrow A$, the effective mass of holes slightly differs from each other and has values of 0.22 and 0.23, respectively. When approaching to $\Gamma$ from $X$, the effective mass of the electrons is nearly the same as ones located around $\Gamma \rightarrow A$  ($\Gamma \rightarrow X$ 0.46, $\Gamma \rightarrow A$ 0.48). 

Work function, a fundamental property of the material, is associated with the ionization energies of atoms in the material. In order to understand the surface properties, work function of $\alpha$-phase is calculated and it is found to be 5.20 eV, which is comparably close to that of MoS$_2$ (5.88 eV)\cite{MYK-graphene}. In addition, it is found that the $\alpha$-phase can be a suitable candidate for solar photocatalysts chemistry since it possesses suitable band edge positions for water-splitting reactions, where the reduction and the oxidation potentials\cite{Kanan} at pH= 7 are -4.03 and -5.26 eV, respectively.

\begin{figure*}
\includegraphics[width=16.5 cm]{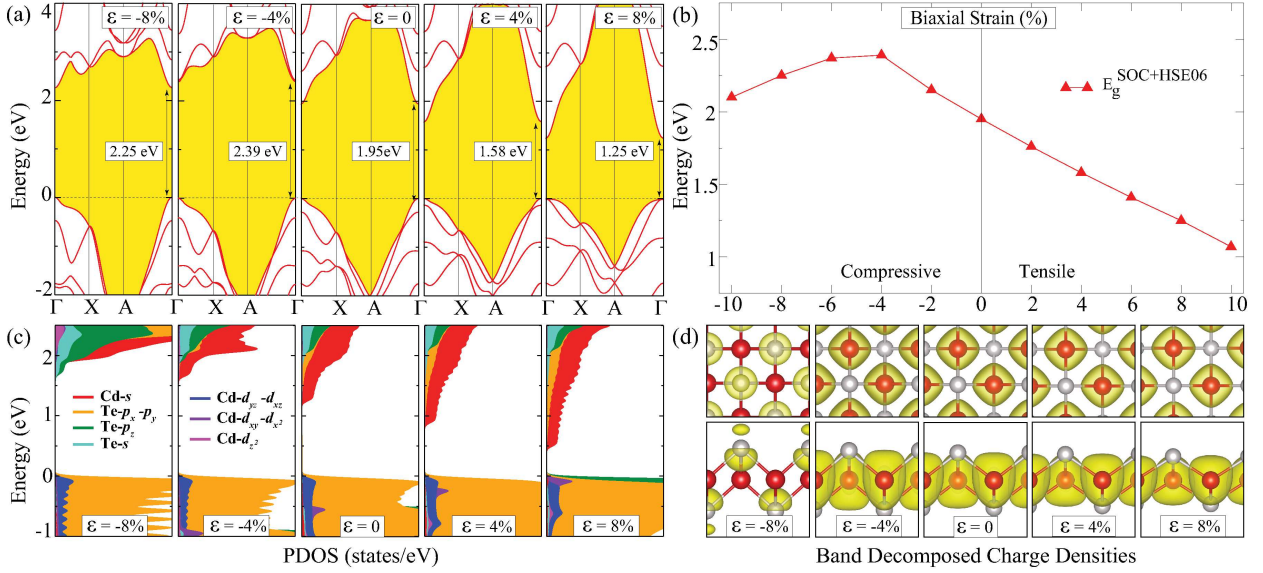}
\caption{\label{strain0}
(Color line) (a) The electronic band diagrams of monolayer $\alpha$-CdTe calculated via SOC-added HSE06 method under compressive and tensile strain. The Fermi level is set to zero. (b) The change in the band gap value with applied biaxial strain. (c) Partial density of states and (d) band decomposed charge densities of monolayer $\alpha$-CdTe under biaxial. }
\end{figure*}

\section{Strain Dependent Properties of $\alpha$-C\MakeLowercase{d}T\MakeLowercase{e}}\label{strain}

The applied strain can cause indirect-to-direct band-gap transition and/or change in the band gap value.\cite{Sahin,Lu-str} In this section, we focus on the strain dependent properties of the monolayer $\alpha$-CdTe and investigate how its mechanical and electronic properties vary with the applied strain. 

Firstly, we apply $\pm 1.5 \%$ strain with a step size of 0.005 and investigate the structural change. It is found that Cd-Te bond increases $1 \%$ with tensile strain and decreases $1 \%$ with compressive strain. This results from the distortion of the charge density with applied strain. Moreover, our results show that thickness of the monolayer $\alpha$-CdTe decreases monotonically from compressive to tensile strain for minimizing the influence of the applied strain.

It is calculated that the $\alpha$-phase has an isotropic in-plane stiffness of 25 N.m$^{-1}$. It appears that compared to stiffness of monolayer MoS$_2$ (122 N.m$^{-1}$) \cite{MYK-graphene} and graphene (340 N.m$^{-1}$\cite{Lee-str}), $\alpha$-CdTe is a quite flexible and soft material. Moreover, Poisson's ratio, that gives the information about perpendicular enlargement in crystal structure when it is stretched in a certain direction, is determined by means of DFT calculations. The Poisson's ratio values of monolayer $\alpha$-CdTe, MoS$_2$ and graphene are 0.28, 0.26\cite{MYK-graphene} and 0.19\cite{MYK-graphene}, respectively. As it is noticed, under applied strain, monolayer MoS$_2$ and $\alpha$-CdTe have the same sensitivity and they are much flexible than graphene.

Furthermore, in order to illustrate trends in band gap for the strained $\alpha$-CdTe (see Fig. \ref{strain0} (a)), we apply biaxial strain in range where the strains are considered to be between $\pm 10 \%$. In each step, lattice parameters were changed by $\pm 2 \%$. During the stretching and shrinking procedure, the structure remains as direct gap semiconductor and the constituent orbitals of VBM remains unchanged. Between the $ -4 \%$ and $ 10 \%$ strain range, Cd-$s$ orbital electrons dominate the conduction band edge (see Fig. \ref{strain0} (c) and (d)). As the structure enlarges, the distance between the $s$-states opens and the interaction between these states weakens; therefore, band gap decreases monotonically as seen in Fig. \ref{strain0} (b). As the structure is compressed more than $-4 \%$, band ordering of the conduction band in energy space changes and the conduction band edge i mostly dominated by Te-$p_z$ orbital electrons. Compressing the structure leads to open the $p_z$-orbitals which results in a decrease in the band gap value. Our results reveal that the band gap of the monolayer $\alpha$-CdTe monolayer can be tunable upon biaxial strain.

\section{Conclusion}\label{conclusion}

In this study, we proposed the thinnest monolayer of CdTe. Structural and vibrational analyses revealed that the structure has same geometry as $\alpha$-PbO and it is found that the structure is dynamically stable. Moreover, we performed Raman-intensity calculations and it was found that the monolayer $\alpha$-CdTe has two Raman-active modes with frequencies. In addition, we also examined its electronic structure by including spin-orbit interaction with HSE06 method. We found that the structure is a direct band gap semiconductor with a gap value of 1.95 eV which falls on the visible range. Moreover, strain-dependent electronic properties of monolayer $\alpha$-CdTe were also studied by applying both compressive and tensile strain. The results showed that the band gap of monolayer $\alpha$-CdTe alters under biaxial strain. Due to the strain tunable moderate direct band gap, monolayer $\alpha$-CdTe is a promising material for the nanoscale optoelectronic applications.  

\section{ Acknowledgments }

Computational resources were provided by TUBITAK ULAKBIM, High Performance and Grid Computing Center (TR-Grid e-Infrastructure). H.S. acknowledges financial support from the TUBITAK under the project number 116C073. H.S. acknowledges support from Bilim Akademisi-The Science Academy, Turkey under the BAGEP program.

\end{document}